\documentclass[sigconf, screen]{acmart}
\renewcommand\footnotetextcopyrightpermission[1]{} % removes footnote with conference information in first column
\usepackage{latexsym}
\usepackage{url}
\usepackage{xcolor}
\definecolor{newcolor}{rgb}{.8,.349,.1}
\usepackage{times}
\usepackage{epsfig}
\usepackage{graphicx}
\usepackage{amsmath}
\usepackage{graphicx}
\usepackage{booktabs}
\usepackage{enumitem}
\usepackage{colortbl}
\usepackage{float}
\usepackage{algpseudocode}
\usepackage[ruled]{algorithm2e}
\usepackage{caption}
\usepackage{subcaption}
\usepackage{multirow}
\usepackage{enumitem}
\usepackage{etoolbox}
\usepackage[normalem]{ulem}
\usepackage{csquotes}
\definecolor{mygray}{gray}{.9}
\definecolor{ForestGreen}{RGB}{34,139,34}

\usepackage{tcolorbox}
\newtcolorbox[list inside=prompt,auto counter,number within=section]{prompt}[1][]{
    colbacktitle=black!60,
    coltitle=white,
    fontupper=\footnotesize,
    boxsep=5pt,
    left=0pt,
    right=0pt,
    top=0pt,
    bottom=0pt,
    boxrule=1pt,
    title={#1},
    #1, % add more args
}
\AtBeginDocument{%
  }

\setcopyright{acmlicensed}
\copyrightyear{2025}
\acmYear{2025}
\acmDOI{XXXXXXX.XXXXXXX}
\begin{document}

\title[MultiFakeVerse Dataset]{Multiverse Through Deepfakes: The MultiFakeVerse Dataset of Person-Centric Visual and Conceptual Manipulations}

\author{Parul Gupta}
\email{parul@monash.edu}
\orcid{0000-0002-4379-1573}
\affiliation{%
  \institution{Monash University}
  \city{Melbourne}
  % \state{Victoria}
  \country{Australia}
}

\author{Shreya Ghosh}
\email{shreya.ghosh@curtin.edu.au}
\orcid{0000-0002-2639-8374}
\affiliation{%
 \institution{Curtin University}
 \city{Perth}
 % \state{Western Australia}
 \country{Australia}}

\author{Tom Gedeon}
\email{tom.gedeon@curtin.edu.au}
\orcid{0000-0001-8356-4909}
\affiliation{%
  \institution{Curtin University}
 \city{Perth}
 % \state{Western Australia}
 \country{Australia}
}

\author{Thanh-Toan Do}
\email{toan.do@monash.edu}
\orcid{0000-0002-6249-0848}
\affiliation{%
  \institution{Monash University}
  \city{Melbourne}
  % \state{Victoria}
  \country{Australia}
}

\author{Abhinav Dhall}
\email{abhinav.dhall@monash.edu}
\orcid{0000-0002-2230-1440}
\affiliation{%
  \institution{Monash University}
  \city{Melbourne}
  % \state{Victoria}
  \country{Australia}
}

\renewcommand{\shortauthors}{Gupta et al.}

\begin{abstract}
  % The dataset contains synthetically manipulated images generated via large language models, where the instructions target modifications to individuals or contextual elements of a scene that influence human perception of importance, intent, or narrative. The focus is on perceptual manipulation rather than identity alteration, bridging face-based deepfakes and broader image semantics.
  
The rapid advancement of GenAI technology over the past few years has significantly contributed towards highly realistic deepfake content generation. Despite ongoing efforts, the research community still lacks a large-scale and reasoning capability driven deepfake benchmark dataset specifically tailored for person-centric object, context and scene manipulations. In this paper, we address this gap by introducing MultiFakeVerse, a large scale person-centric deepfake dataset, comprising 845,286 images generated through manipulation suggestions and image manipulations both derived from vision-language models (VLM). The VLM instructions were specifically targeted towards modifications to individuals or contextual elements of a scene that influence human perception of importance, intent, or narrative. This VLM-driven approach enables semantic, context-aware alterations such as modifying actions, scenes, and human-object interactions rather than synthetic or low-level identity swaps and region-specific edits that are common in existing datasets. Our experiments reveal that current state-of-the-art deepfake detection models and human observers struggle to detect these subtle yet meaningful manipulations. The code and dataset are available on \href{https://github.com/Parul-Gupta/MultiFakeVerse}{GitHub}.

\end{abstract}


% \ccsdesc[500]{Computing methodologies~Computer vision}
% \ccsdesc[500]{Security and privacy~Social aspects of security and privacy}
% \ccsdesc[300]{Security and privacy~Usability in security and privacy}

% A "teaser" image appears between the author and affiliation
% information and the body of the document, and typically spans the
% page.
\begin{teaserfigure}
  \includegraphics[width=\textwidth]{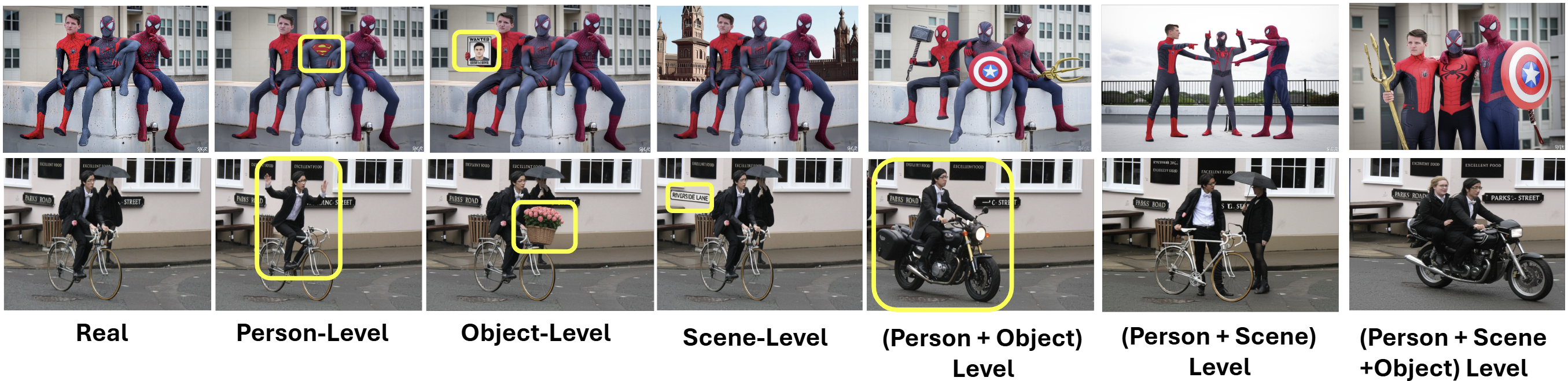}
  \caption{MultiFakeVerse. A brief overview of the proposed dataset. Here, we introduce subtle and profound person-centric deepfakes covering \textit{person-level}, \textit{object-level}, \textit{scene-level}, \textit{(person+object)-level}, \textit{(person+scene)-level} manipulations. Image best viewed in color.}
  \Description{}
  \label{fig:teaser}
\end{teaserfigure}

\received{20 February 2007}
\received[revised]{12 March 2009}
\received[accepted]{5 June 2009}

\maketitle

\section{Introduction}
\label{sec:intro}
The last decade has witnessed an unprecedented surge in the capabilities of content generation technologies powered by deep learning. Models trained on massive and diverse datasets now possess the ability to generate content that is often indistinguishable from authentic data. This advancement spans across multiple modalities, including text~\cite{brownLanguage2020, touvronLlama2023a}, images~\cite{huang2024sida}, videos~\cite{ge2022long}, and audio~\cite{shen2023naturalspeech}. Such generative models ranging from large language models (LLMs) to generative adversarial networks (GANs) and diffusion models have transformed the creative and information landscapes. However, while these models present numerous opportunities in entertainment, education, accessibility and design, they also present a potential negative societal impact.

In recent years, numerous data sets have been proposed to facilitate the training and evaluation of deepfake detectors~\cite{cai2023glitch,caiYou2022,caiavdeepfake1m,wang2020cnn}. These datasets often span different modalities and types of manipulations. For instance, visual-only datasets focus on face-swapping, facial reenactment, or expression synthesis~\cite{narayanDFPlatter2023}; while audio-only datasets target voice cloning or speech synthesis~\cite{casanovaSCGlowTTS2021}; and audiovisual datasets integrate both, enabling the study of multimodal manipulations~\cite{caiavdeepfake1m}. These resources have proven invaluable for benchmarking detection algorithms, enabling the deepfake research community to systematically compare methods and track progress. However, despite these efforts, one of the most significant limitations is in terms of human-centric factors present in the image, specially from human-object and human scene interaction reasoning perspective (as shown in Fig.~\ref{fig:teaser}).

To this end, we propose \textit{MultiFakeVerse} dataset, a collection of synthetically manipulated images created using large Vision-language models (VLMs), designed to explore perceptual modifications within visual content. Unlike traditional deepfake datasets that primarily focus on altering facial identities or facial expressions, \textit{MultiFakeVerse} emphasizes targeted modifications to individuals or contextual elements within a scene to influence human perception. These alterations are carefully crafted to change the narrative, importance, or intent perceived by viewers, often without modifying the core identity of subjects. This approach bridges the gap between face-based deepfakes and broader image semantics by focusing on semantic and contextual manipulations that can impact how an image's story or message is interpreted. Such modifications can include changing background elements, repositioning objects, or altering actions to shift the scene’s overall meaning or emphasis. The purpose of \textit{MultiFakeVerse} is to challenge detection methods and deepen understanding of perceptual manipulation, as these subtle yet impactful changes can be more challenging to identify than overt facial swaps. By concentrating on perceptual cues rather than identity alterations, the dataset provides a valuable resource for advancing research in robust detection algorithms against manipulative visual content. The main contributions of the paper are:

\begin{table}[t]
\centering
\caption{\textbf{Details for publicly available deepfake datasets in a chronologically ascending order.} }
% \vspace{-3mm}
\label{tab:datasets}
\scalebox{0.8}{
\begin{tabular}{l||c|c|c|c}
\toprule[0.4mm]
\rowcolor{mygray}\textbf{Dataset} & \textbf{Year} & \multicolumn{2}{|c|}{\textbf{Manipulation}} & \textbf{\#Total} \\
\rowcolor{mygray}&  & \textbf{Method} & \textbf{Content} & \\
\hline\hline
% DF-TIMIT~\cite{korshunov2018deepfaketimit} & 2018  & Face swapping & Face & 960 \\
% UADFV~\cite{yangExposing2019} & 2019  & Face swapping & Face & 98  \\
% FaceForensics++~\cite{rosslerFaceForensics2019} & 2019  & Face reenactment & Face & 5,000 \\
% Google DFD~\cite{nickContributing2019} & 2019  & Face swapping & Face  & 3,431 \\
% DFDC~\cite{dolhanskyDeepFake2020} & 2020  & Face swapping & Face & 128,154 \\
% DeeperForensics~\cite{jiangDeeperForensics12020} & 2020  & Face swapping & Face & 60,000 \\
% Celeb-DF~\cite{liCelebDF2020} & 2020  & Face swapping & Face  & 6,229 \\
% WildDeepfake~\cite{ziWildDeepfake2020} & 2020  & Face swapping & - & 7,314 \\
DFFD~\cite{stehouwer2019detection}& 2019  & GAN & Face   &  299,039 \\
FakeSpotter~\cite{wang2019fakespotter}& 2020  & GAN & Face   &  11,000 \\
FFHQ~\cite{karras2019style}& 2020  & GAN & Face   &  70,000 \\
 
CNNSpot~\cite{wang2020cnn}& 2020  & GAN & Face \& Objects  &  724,000 \\
\multirow{2}{*}{IMD2020~\cite{novozamsky2020imd2020}} & \multirow{2}{*}{2020}  & GAN  & Face, Objects &  \multirow{2}{*}{105,000} \\
&   &  Inpainting & \& Scene &   \\
% FFIW$_{10K}$~\cite{zhouFace2021} & 2021  & Face swapping & Face & 20,000 \\
% KoDF~\cite{kwonKoDF2021} & 2021  & Face swapping & Face & 237,942 \\
% FakeAVCeleb~\cite{khalidFakeAVCeleb2021} & 2021  & Face reenactment & Face & 25,500$+$ \\
% ForgeryNet~\cite{heForgeryNet2021} & 2021  & Face reenactment & Face & 221,247 \\
% LAV-DF~\cite{caiYou2022} & 2022  & Face reenactment & Face & 136,304 \\
% DF-Platter~\cite{narayanDFPlatter2023} & 2023  & Face Swapping & Face & 265,756 \\
OHImg~\cite{may2023comprehensive}& 2023  & Diffusion & Overhead  &  13,150\\
ArtiFact~\cite{rahman2023artifact} & 2023  & Diffusion \& GAN & Objects  & 2,496,738 \\
AIGCD~\cite{zhong2023rich} & 2023  & GAN & Objects & 1,440,000 \\
GenImage~\cite{zhu2023genimage} & 2023  & Diffusion \& GAN & Objects & 2,681,167 \\
% AV-DeepFake1M~\cite{caiavdeepfake1m} & 2024  &  & Face & 1,146,760 \\
CiFAKE~\cite{bird2024cifake}& 2024  & CIFAR & Objects  &  120,000\\
M3Dsynth~\cite{zingarini2024m3dsynth}& 2024  & Diffusion & Medical  & 8,577 \\

SemiTruths~\cite{pal2024semi} & 2024  & Diffusion & Face, Objects & 1,500,300 \\
SIDA~\cite{huang2024sida} & 2024  & Diffusion & Objects   &  300,000\\
\hline
\multirow{2}{*}{\textbf{MultiFakeVerse}} & \multirow{2}{*}{2025}  & Vision-Language &  Face, Persons,  &  \multirow{2}{*}{845,286}\\
&&Models&Objects, Scene, Text&\\
\bottomrule[0.4mm]
\end{tabular}}
% \vspace{-5mm}
\end{table}

\begin{itemize}[topsep=1pt,itemsep=0pt,partopsep=1ex,parsep=1ex,leftmargin=*]
    \item We propose \textit{MultiFakeVerse}, a large-scale reasoning-driven human-centric deepfake dataset for the task of deepfake detection. The proposed dataset mainly explore novel image level perceptual modifications within visual content in terms of \textit{human gestures}, \textit{human-object interaction} and \textit{human-scene interaction}.
    \item We propose a novel data generation pipeline employing novel \textit{LLM-based image perception} manipulation strategies which include \textit{human gestures}, \textit{human-object interaction} and \textit{human-scene interaction} and incorporating the state-of-the-art (SOTA) in image generation.
    \item We perform comprehensive analysis and benchmark the proposed dataset with SOTA deepfake detection methods.
\end{itemize}

\section{Related Work}

\subsubsection{\textbf{Deepfake Datasets}} The current landscape of image-based deepfake datasets is shown in Table~\ref{tab:datasets}. Most existing deepfake datasets focus primarily on person-level changes, where manipulations are confined to a person's face or facial expressions. A prominent example is DFFD~\cite{stehouwer2019detection}, which contains 300K samples, which are generated and/or edited facial images with GAN methods. Other prominent datasets, which have been commonly used are FFHQ~\cite{karras2019style} and FakeSpotter ~\cite{wang2019fakespotter}. 
% Research in face focused deepfakes detection has also gained from video-level deepfakes datasets such as DFDC~\cite{dolhansky2020deepfakedetectionchallengedfdc}, CelebAVFakes~\cite{liCelebDF2020}, DF-Platter~\cite{narayanDFPlatter2023} and AV-Deepfakes-1M~\cite{caiavdeepfake1m}.
Moving beyond faces, OHImg~\cite{may2023comprehensive} introduced a deepfake dataset focused on overhead aerial imagery, while M3Dsynth~\cite{zingarini2024m3dsynth} addressed medical image based manipulations. More recently, Huang et al. proposed the SIDA dataset~\cite{huang2024sida}, which leverages vision-language models to identify objects in images and captions, replacing them via inpainting with Stable Diffusion (e.g. cat $\rightarrow$ dog). However, these datasets primarily focus on object- and scene-level manipulations and do not address person-level edits that alter the perceived meaning of an image. 

\begin{figure*}[t]
    \centering
    \begin{subfigure}[b]{0.75\linewidth}
        \centering
        \includegraphics[width=\linewidth]{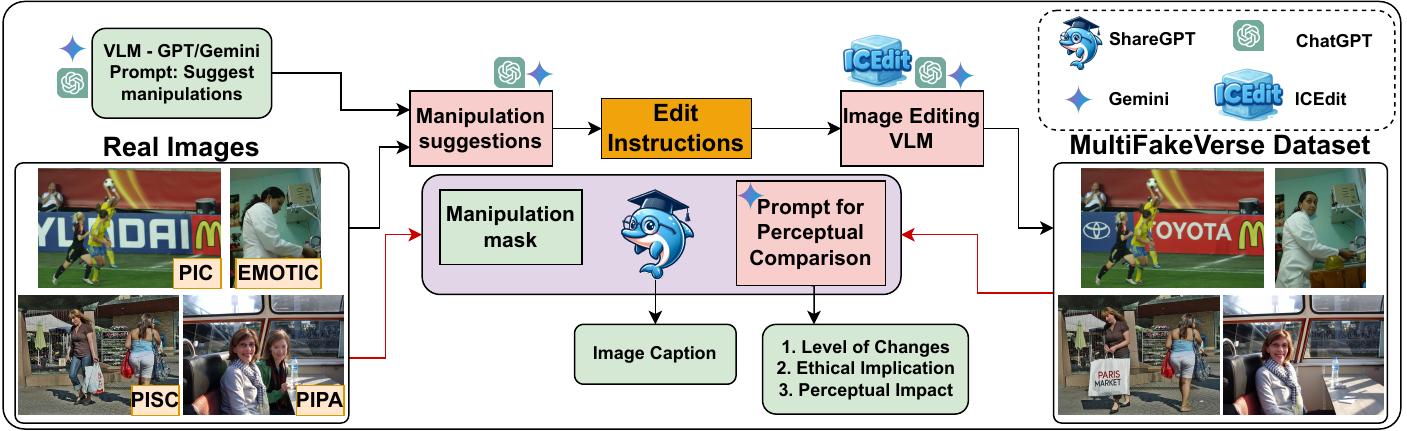}
        \caption{Dataset generation pipeline.}
        \label{fig:gen_pipeline}
    \end{subfigure}
    \hfill
    \begin{subfigure}[b]{0.23\linewidth}
        \centering
        \includegraphics[width=\linewidth]{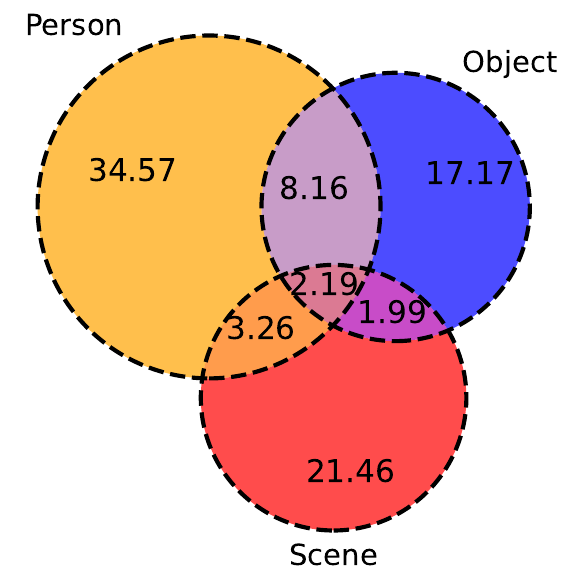}
        \caption{Different Manipulation levels.}
        \label{fig:venn_levels}
    \end{subfigure}
    
    \caption{(a) Depicts the MultiFakeVerse dataset generation pipeline. For details please see Section \ref{sec:Dataset}. (b) Venn Diagram depicting the semantic categorical-levels of manipulations in the MultiFakeVerse dataset. }
    \Description{Creating Semantic context-aware deepfakes.} 
    % \vspace{-2mm}
    \label{fig:dataset_stats}
\end{figure*}

Another interesting recent work, SemiTruths ~\cite{pal2024semi}, proposed a large-scale dataset comprising varying levels of manipulation, focusing on objects and scenes. Both inpainting (Stable Diffusion) and prompt-based (LLAMA-7B \cite{touvron2023llama}) methods are used for introducing image manipulations. The focus is mainly on objects and scenes with the change criterion being change in perceptual meaning of the image at varying levels. In contrast, our work focuses exclusively on people in images. This focus is critical: studies show that deepfakes targeting real individuals are shared and viewed up to six times more often than generic content~\cite{vosoughi2018spread}, and that over 96\% of deepfakes online are non-consensual pornography~\cite{wired2019deepfakes}. This underscores the urgent societal impact of person-centered manipulations, an area where current datasets are lacking.

Existing datasets predominantly focus on within-face manipulations or object-scene perturbations, which do not capture the complexity of real-world deepfakes. However, person-centered deepfakes often involve diverse manipulations such as full-body edits, co-occurring people, context shifts, object edits and scene re-compositions, which pose unique challenges for detection methods. Therefore, a dataset focusing on these person-centric manipulations is essential for advancing deepfakes detection systems.

\subsubsection{\textbf{Deepfake Detection}} Deepfake detection techniques are commonly framed as classification problems within a data-centric framework~\cite{sharma2024systematic}. These approaches often rely on a range of neural network architectures, notably Convolutional Neural Networks (CNNs)~\cite{nguyen2019deep} and Transformers~\cite{wang2024timely}, to identify unique artifacts introduced during image based manipulation. A number of deepfake detection methods~\cite{miao2023msccnet, huang2024sida} have expanded beyond simple binary classification by developing datasets with pixel-level annotations of tampered regions, thereby supporting both detection and localization tasks. However, such resources are predominantly centered on facial forgeries~\cite{miao2023msccnet}, with limited availability of datasets targeting non-facial manipulations, and even fewer large-scale, publicly accessible datasets that reflect the diversity of scene content. To bridge these gaps, our work introduces a comprehensive dataset featuring a wide range of manipulations beyond facial edits, with a particular emphasis on realistic image-based manipulation and localization.

\subsubsection{\textbf{LLM Powered Reasoning for Generation}} The emergence of diffusion models has significantly advanced the field of image editing/manipulations~\cite{wang2025image}. Recent progress in image inpainting, under both text-guided and unguided settings, has enabled more precise and high-quality manipulations~\cite{wang2025image}. Text-conditioned inpainting~\cite{xie2023smartbrush} allows for detailed control over edits using natural language, while unconditioned methods offer strong baseline capabilities without explicit guidance. Traditionally, inpainting techniques~\cite{huang2024sida} rely on the use of masks to specify regions for modification. In contrast, prompt-based image editing~\cite{touvron2023llama} allows for targeted alterations driven solely by text inputs, eliminating the need for explicit masks. Frameworks such as LANCE~\cite{prabhu2023lance} and InstructPix2Pix~\cite{brooks2022instructpix2pix} harness this approach to automate image perturbation pipelines. LANCE~\cite{prabhu2023lance}, in particular, combines large language models (LLMs) with image captioning techniques to facilitate diverse, human-free image editing workflows. Expanding on this line of work, our method uses GPT-Image-1 ~\cite{openai2024gpt4o}, Gemini-2.0-Flash-Image-Generation
\cite{anil2023gemini} and ICEdit \cite{icedit_2025} to support image manipulation with a wider spectrum of perturbation intensities, guided by definitions of semantic change. To achieve this, we integrate models like Gemini-2.0-Flash \cite{anil2023gemini}, ChatGPT-4o-latest~\cite{openai2024gpt4o} using prompt-based editing to generate precise, semantically meaningful image modifications. Multimodal models like Gemini, ChatGPT-4o, and Icedit leverage both text and image inputs to enable precise, context-aware edits and have a natural advantage over text-driven models like LANCE and InstructPix2Pix in fine-grained and semantically consistent image manipulation tasks. While text-driven models are effective for stylistic transformations, multimodal approaches offer broader capabilities and greater control over localized edits.
\section{MultiFakeVerse Dataset} \label{sec:Dataset}
To create the MultiFakeVerse dataset for person-centric manipulations, we use publicly available real image datasets involving humans in diverse environments. EMOTIC~\cite{kosti2019context}, PISC~\cite{li2017dual}, PIPA~\cite{pipa_dataset} and PIC 2.0~\cite{pic_dataset}. Altogether, we use $86,952$ images from these datasets to create a total of $758, 041$ fake images. This huge scale positions our proposed dataset as the most comprehensive reasoning-guided deepfake benchmark. The dataset generation details are below:
\subsection{Data Generation Pipeline}

\begin{prompt}[title={Prompt \thetcbcounter: VLM based Image Perception Manipulation}]
\textbf{Human:} \{INPUT image to VLM\} Given the attached image, identify the most important person in this image, and suggest minimal modifications to the image to obtain each of the following effects:\\
(1) the person you identified appears naive\\
(2) the person you identified appears nonchalant\\
(3) the person you identified appears proud\\
(4) the person you identified appears remorseful\\
(5) the person you identified appears inexperienced\\
(6) some factual information depicted in the image changes\\
The possible change targets for the modifications are: the objects or text or humans in the image. When suggesting changes to text in the image, be specific about what text is to be replaced and what text should be added instead.
Give output as a valid JSON string in the following format:

\{`Most Important Person':<referring expression for most important person>\\
`Effects':[\{`Effect':<effect>, `Change Target': <change target>, `Explanation': [<referring expression for the change target>, <edit\_instruction>]\}]\}.\\
Do not include any other information in the response.

% \textbf{AI:} \{EXAMPLE OUTPUT 1\} (\sg{See Figure******})

% $...... $ % it means there are multiple examples provided.
\textbf{Human:} \{INPUT image to VLM\} In this image, change \{target\} \{edit\_instruction\}. Do not change anything else in the image.

% \textbf{AI:} \{EXAMPLE OUTPUT 2\} (\sg{See Figure******})
\end{prompt}

\subsubsection{\textbf{Curation of Person-centric edit instructions}}
As the first step in our approach, we leverage the extensive visual understanding capability of foundational VLMs (particularly Gemini-2.0-Flash and ChatGPT-4o-latest) to obtain six suitable edits for each image, aimed towards altering the viewer's perception of the most important person in the image. To this end, we ask the VLM for \textit{minimal} edits that would make the person appear too \textit{naive}, \textit{proud}, \textit{remorseful}, \textit{inexperienced}, \textit{nonchalant}, or some factual information depicted in the image would change. To ensure that the edits can be used accurately in the further stages, the output includes the \textit{referring expression} of the target of each modification along with the edit instruction. Note that \textit{referring expression} is a widely explored domain in the community, which means a phrase which can disambiguate the target in an image, e.g. for an image having two men sitting on a desk, one talking on the phone and the other looking through documents, a suitable referring expression of the later would be \textit{the man on the left holding a piece of paper}.

\subsubsection{\textbf{VLM based Image perception Manipulation}}
In this step, for each of the $<$referring expression, edit instruction$>$ pair, we prompt a VLM to perform the edit on the image, while ensuring that nothing else changes. We experiment with three different VLMs here - GPT-Image-1, Gemini-2.0-Flash-Image-Generation and ICEdit~\cite{icedit_2025}. After observing ~22K generated images, Gemini emerged as the VLM for highest quality manipulations, as its edits are the most coherent with the rest of the scene with no changes to the untargeted regions of the input image.The ICEdit model often results in easily identifiable fakes with significant artifacts, whereas GPT-image-1 tends to edit in a few cases, untargeted regions due to constraints in the output aspect ratios (current options are: $1024\times1024$, $1536\times1024$, $1024\times1536$).

\subsection{Analyzing the Manipulated images}
\label{subsec:manipulation_analysis}
To further enrich our dataset, we perform the following analyses on the generated images:

\begin{table}[b]
\centering
% \vspace{-4mm}
\caption{\textbf{Number of images in MultiFakeVerse.}}
% \vspace{-3mm}
\scalebox{0.85}{
\begin{tabular}{l|ccc|c}
\toprule[0.4mm]
\rowcolor{mygray} \textbf{Subset} & \textbf{Source} & \textbf{\#Real Images} & \textbf{\#Fake Images} & \textbf{\#Images} \\ \hline \hline
\multirow{4}{*}{Train} & Emotic & 15976 & 176052 & \multirow{4}{*}{592,349}  \\
& PIC 2.0 & 8565  & 54320 & \\
& PIPA & 20163  & 115730 &  \\
& PISC & 16677 & 184866 &  \\ \hline
\multirow{4}{*}{Validation} & Emotic & 2226 & 25132 & \multirow{4}{*}{84,309} \\
& PIC 2.0 & 1203  & 7450  &  \\
& PIPA & 2864 & 16564 &  \\
& PISC & 2327  & 26543  &  \\ \hline
\multirow{4}{*}{Test} & Emotic & 4453 & 50315 &  \multirow{4}{*}{168,628} \\ 
& PIC 2.0 & 2406 & 15383 &  \\
& PIPA & 5730   & 32961 &  \\
& PISC & 4655  & 52725  &  \\ \hline \hline
\multirow{4}{*}{Overall} & Emotic & 22655 & 251499  &  \multirow{4}{*}{845,286}\\
& PIC 2.0 & 12174 & 77153 &  \\
& PIPA & 28757 & 165255 &  \\ 
& PISC & 23659 & 264134 &  \\ 
\bottomrule[0.4mm]
\end{tabular}}
\label{tab:dataset_stats}
% \vspace{-4mm}
\end{table}

\subsubsection{\textbf{Ratio of edited area}} To obtain a measure of the spatial degree of modification, we calculate the mean squared difference between the pixel values of the original and altered images, threshold the difference to remove noise, and perform connected component analysis to obtain the masks of the edited regions. The ratio of edited area to the total image area is then plotted over the entire dataset, shown in Figure \ref{fig:area_dist}, which displays a distribution spread out from 0 to 0.8, thus signifying the wide variety of edits present in our dataset.

\subsubsection{\textbf{Image Captioning}} We use ShareGPT4V ~\cite{chen2024sharegpt4v} VLM model to obtain rich captions of both the original as well as tampered images. These are then used to calculate the directional similarity~\cite{directional_clip} between the real versus edited images and their corresponding captions as follows: ($CLIP_{Original Image}$ - $CLIP_{Tampered Image}$) $\cdot$ ($CLIP_{Original Image Caption}$ - $CLIP_{Tampered Image Caption}$). Particularly, we use Long-CLIP~\cite{zhang2024longclip} to calculate the image and text embeddings. From this analysis, we find that the maximum semantic change is observed in the object category, where the modification is around or on the person.

The next set of analyses are performed through the Gemini-2.0-Flash VLM to compare the real and manipulated images and measure the perceptual changes for different concepts. 
\subsubsection{\textbf{Level of manipulation performed}} We categorize the manipulations into Person-level, Object-level and Scene-level depending upon whether the manipulation target is a person (facial expression, identity, gaze, pose, or clothing), an object connected to a person (i.e. an object in the foreground - appearance change, addition, removal etc.) or some component in the overall scene/background of the image respectively. Since an image can have multiple levels of changes depending upon the edit instruction, we display the distribution of the these alteration levels in our dataset through a Venn diagram in Figure \ref{fig:venn_levels}. The distribution shows that the different levels of manipulations are widely distributed in our dataset with around one-third of the edits being person-only, around one-fifth being scene-only and one-sixth being object-only.

\begin{prompt}[title={Prompt \thetcbcounter: VLM based Image Perception Analysis}] \label{prompt:vlm_image_perception}
% \textbf{System:} ChatGPT and Gemini
\textbf{Human:} \{EXAMPLE INPUT image and Manipulated image\} Compare these two images: the first is the original (Image A) and the second is its manipulated version (Image B). For each of the following categories, describe the changes observed, if any, and their potential perceptual impact on a human viewer. Finally, classify the changes in Image B into person-level, person-object level or person-scene level, depending upon whether a person has been changed, an object connected to the person has been changed or a component of the scene away from the person has been changed respectively. Note that a manipulated image can have multiple types of changes, so output multiple labels for such image. Strictly use the following structure in your response, and do not add anything else:\\
1. Emotion/Affect:
- Image A: [Describe emotion, mood, facial expression, atmosphere]
- Image B: [Describe changes]
- Perceptual Impact: [How might this affect viewer emotion or interpretation?]

2. Identity \& Character Perception:
- Image A: [Describe apparent age, gender, trustworthiness, etc.]
- Image B: [Describe changes]
- Perceptual Impact: [Does this shift how the person is perceived?]

3. Social Signals \& Status:
- Image A: [Describe clothing, posture, social roles, proximity]
- Image B: [Describe changes]
- Perceptual Impact: [Do power dynamics or relationships change?]

4. Scene Context \& Narrative:
- Image A: [Describe implied story or setting]
- Image B: [Describe changes]
- Perceptual Impact: [Does the story or situation change?]

5. Manipulation Intent:
- Description: [What might be the intent behind the manipulation?]
- Perceptual Impact: [Does the edit appear deceptive, persuasive, aesthetic, etc.?]

6. Ethical Implications:
- Description: [Could the manipulation mislead or cause harm?]
- Assessment: [Mild / Moderate / Severe ethical concern]

7. Type of changes:
 - [Person level / Person-Object level / Person-Scene level]
% \textbf{AI:} \{EXAMPLE OUTPUT 1\} (\sg{See Figure******})
\end{prompt}

\subsubsection{\textbf{Perceptual Impact of the edit on the viewer}}
\begin{figure}[t]
    \centering
    % Row 1
    \begin{subfigure}[b]{0.48\linewidth}
        \centering
        \includegraphics[width=\linewidth]{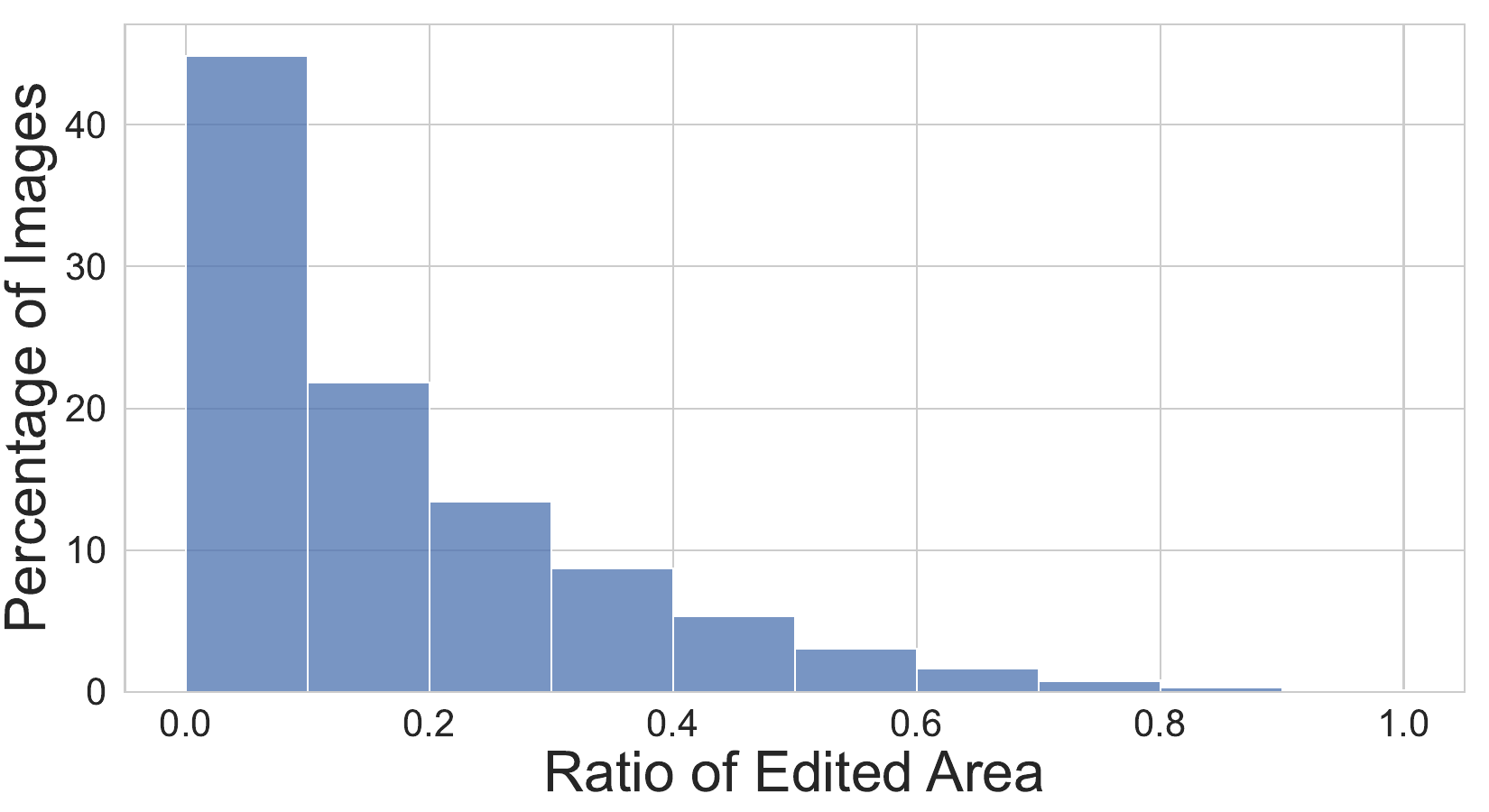}
        \caption{Ratio of Edited Regions.}
        \label{fig:area_dist}
    \end{subfigure}
    \hfill
    \begin{subfigure}[b]{0.48\linewidth}
        \centering
        \includegraphics[width=\linewidth]{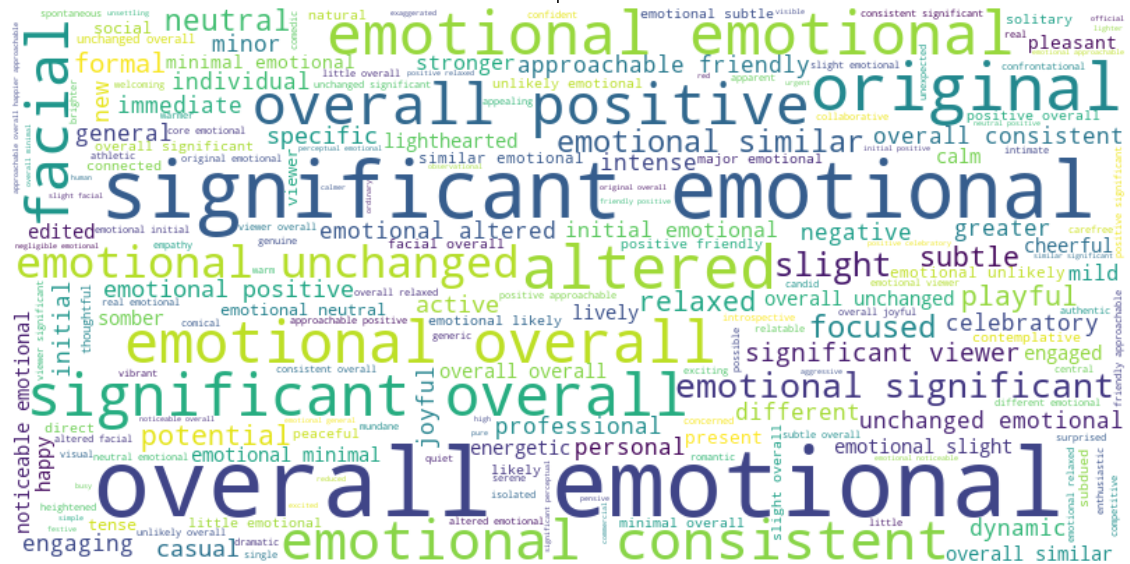}
        \caption{Emotion Perception.}
        \label{fig:emotional_impact}
    \end{subfigure}

    \vspace{0.1cm}

    % Row 2
    \begin{subfigure}[b]{0.48\linewidth}
        \centering
        \includegraphics[width=\linewidth]{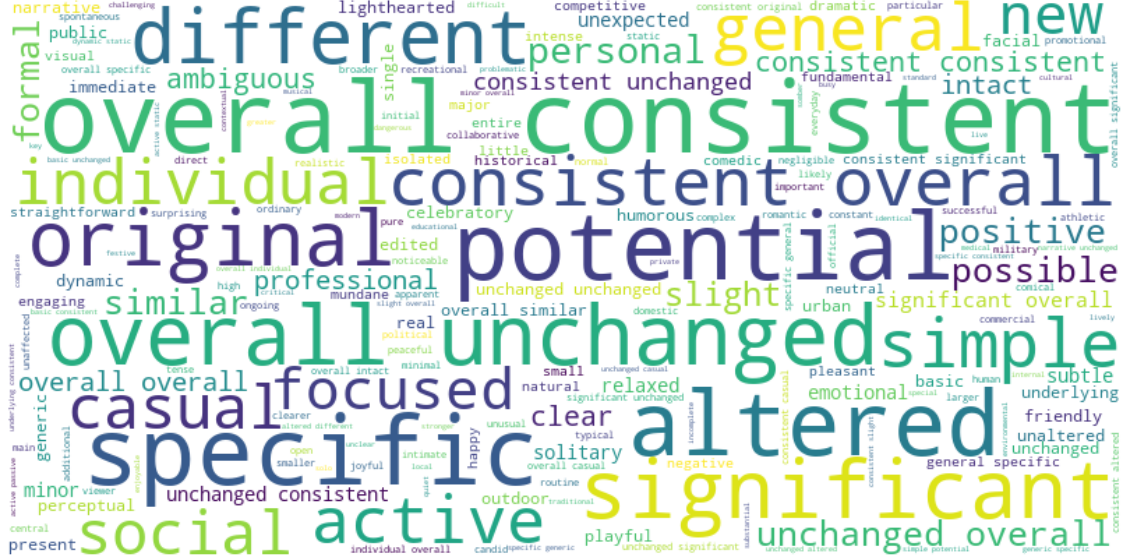}
        \caption{Scene Context and Narrative shift.}
        \label{fig:narrative_shift}
    \end{subfigure}
    \hfill
    \begin{subfigure}[b]{0.48\linewidth}
        \centering
        \includegraphics[width=\linewidth]{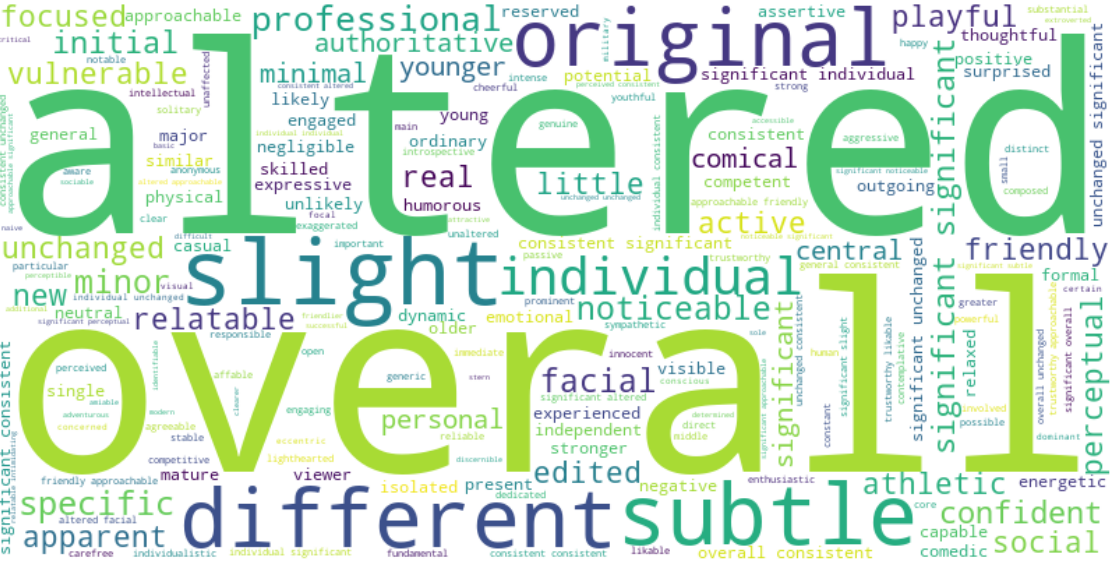}
        \caption{Perceived identity of depicted individual.}
        \label{fig:identity_impact}
    \end{subfigure}

    \vspace{0.1cm}

    % Row 3
    \begin{subfigure}[b]{0.48\linewidth}
        \centering
        \includegraphics[width=\linewidth]{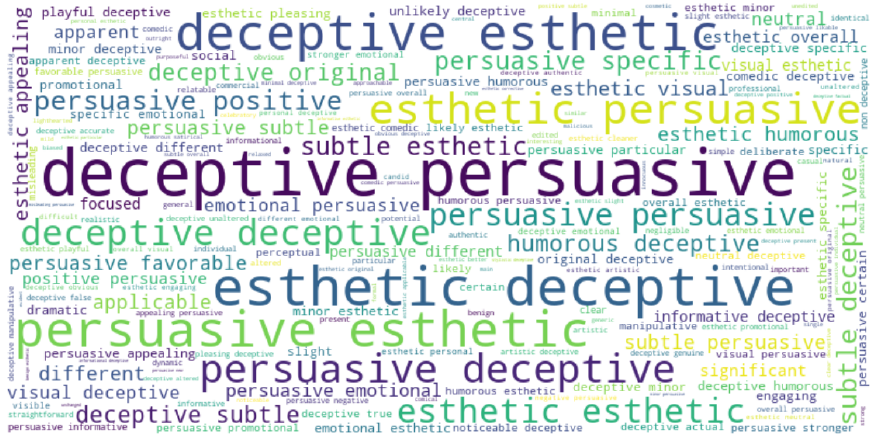}
        \caption{Intent of Manipulations.}
        \label{fig:manipulation_intent}
    \end{subfigure}
    \hfill
    \begin{subfigure}[b]{0.48\linewidth}
        \centering
        \includegraphics[width=\linewidth]{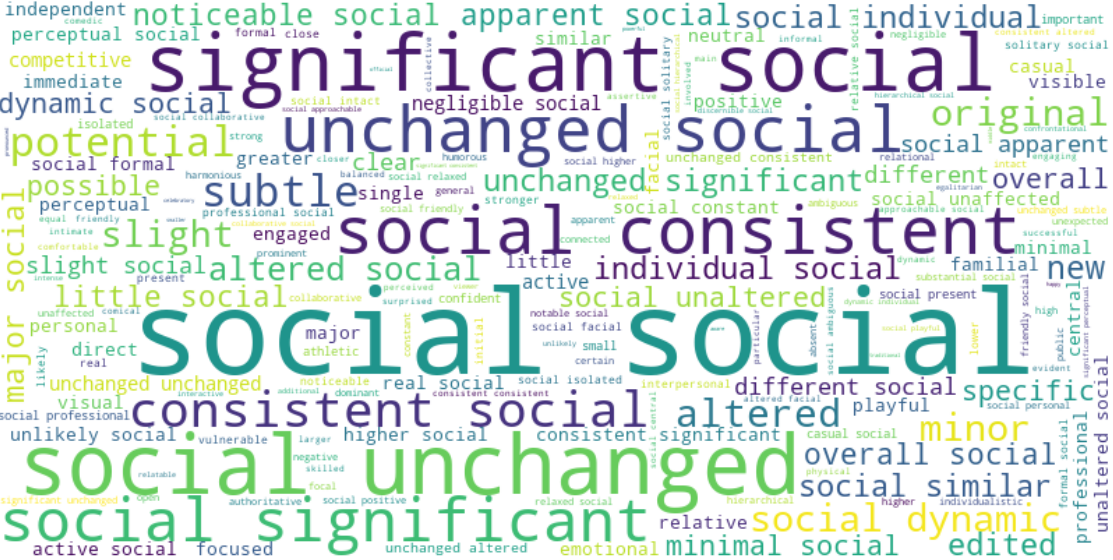}
        \caption{Power dynamics and relationships.}
        \label{fig:social_status_impact}
    \end{subfigure}
    % \vspace{-1mm}
    \caption{The visualizations illustrate the characteristics and impact of fake images in the MultiFakeVerse dataset. (a) shows the distribution of the ratio of pixels edited in fake images. (b–f) display word maps highlighting the perceptual changes observed by viewers, derived using VLM analysis. These word maps capture how image manipulations affect the following attributes: perceived emotion, scene context and narrative shift, perceived identity, intent of manipulation, and impact on power dynamics and relationships.}
   % \vspace{-4mm}
    \label{fig:word_Clouds}
\end{figure}

\begin{figure*}[t]
    \centering
    \vspace{-2mm}
    \includegraphics[width=0.94\linewidth]{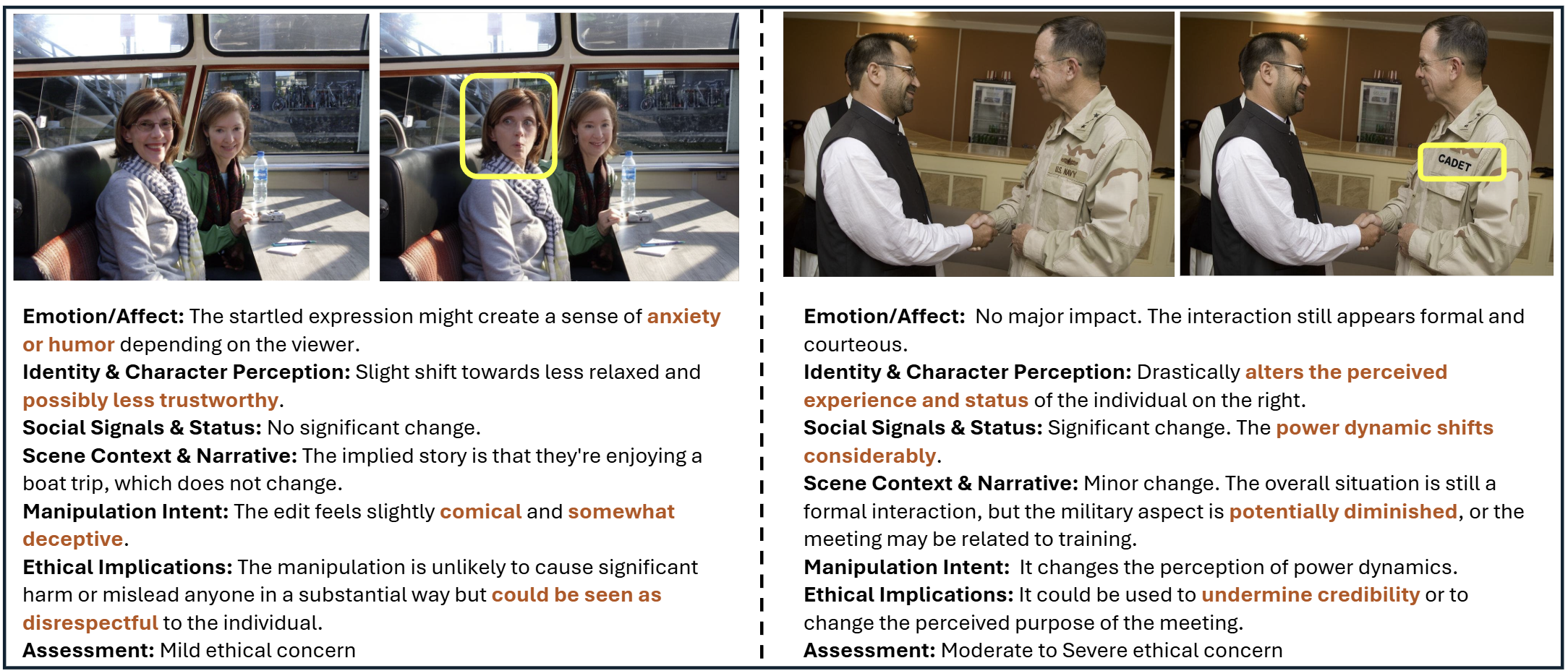}
    \caption{Analyzing the perceptual impact of manipulations in images. The edited regions are highlighted by yellow boxes. The analysis covers changes in attributes such as perceived emotion, identity, and ethical implications.}
    \vspace{-2mm}
    \label{fig:qualitative}
\end{figure*}

To obtain a measure of the change in the viewer perception as a result of the image tamperings, we consider the impact on the following six perceptual factors: (A) the viewer's \textbf{emotion} or interpretation (B) the \textbf{perceived identity or character of the person} in the image (C) the \textbf{perceived power dynamics/relationships} (D) the \textbf{perceived scene narrative} (E) the possible \textbf{intent of manipulation}, and (F) the \textbf{ethical implications} of the tamperings. We prompt the Gemini-2.0-Flash VLM to describe these impacts, whose instruction details are shown in Prompt  \ref{prompt:vlm_image_perception}. The word clouds of the changes based on responses obtained from the VLM are shown in Figure \ref{fig:word_Clouds}. The first word map, corresponding to the changes in emotion perception (Figure \ref{fig:emotional_impact}) features words such as \textit{significant, joyful, engaging, approachable, facial} suggesting that our semantic modifications subtly influence the emotional perception. In case of scene context and narrative shifts (Fig. \ref{fig:narrative_shift}), the word cloud contains terms such as \textit{focused, professional, potential, different}, indicating plausible alterations to the conveyed story or setting due to the corresponding edits. Fig. \ref{fig:identity_impact}, which illustrates the word map for shift in the perceived identity of depicted persons includes terms such as\textit{playful, younger, vulnerable} implying that our manipulations are able to influence the individual's identity. The word map representing the intent behind the manipulations in Fig. \ref{fig:manipulation_intent} shows \textit{deceptive}, \textit{persuasive} and \textit{aesthetic} as the most prominent words, reflecting a tendency for the manipulations to be viewed as intentional acts of persuasion or deception. According to the word map in Fig. \ref{fig:social_status_impact}, though the power dynamics and relationships remain stable for many cases, the presence of words like \textit{altered, dynamic, formal, competitive} highlights that such dynamics do shift noticeably in a subset of the images. In Figure \ref{fig:qualitative}, we show some sample responses obtained for two such edits, with the different perceptual impacts highlighted. The VLM responses show that around 81\% of the edits have mild ethical implications, while 14.2\% have moderate and 0.3\% of the edits have severe ethical implications in our dataset. 

\subsubsection{\textbf{Quality of generations}} To evaluate the visual quality of MultiFakeVerse, we used the peak signal-to-noise ratio (PSNR), structural similarity index (SSIM)~\cite{wangImage2004} and Fréchet inception distance (FID)~\cite{heusel2017gans} metrics as shown in Table~\ref{tab:visual_quality}. SSIM value 0 indicates no similarity between the compared images, while 1 indicates perfect similarity, therefore, our value of $0.5774$ is indicative of the targeted edits which ensure that the untargeted regions remain as close as possible to the original image. Similarly, a low FID value of 3.30 indicates that the quality and diversity of the generated fakes is excellent. A high PSNR value of 66.30 dB is indicative of good image quality.

\begin{table}[t]
\centering
\caption{\textbf{Visual quality of MultiFakeVerse.} \textmd{Quality of the generated images in terms of PSNR, SSIM and FID.}}
% \vspace{-2mm}
\scalebox{0.85}{
\begin{tabular}{l|ccc}
\toprule[0.4mm]
\rowcolor{mygray} \textbf{Dataset} & \textbf{PSNR($\uparrow$)} & \textbf{SSIM($\uparrow$)} & \textbf{FID($\downarrow$)} \\ \hline \hline
% SIDA~\cite{} &  &  &  \\ 
% \sg{if anything} & - & - &  \\ \hline
% MultiFakeVerse (Train) &    &    &    \\
% MultiFakeVerse (Validation) &    &    &   \\
% MultiFakeVerse (Test) &    &    &   \\
MultiFakeVerse&  \textbf{66.30}  &  \textbf{0.5774}  & \textbf{3.30} \\
\bottomrule[0.4mm]
\end{tabular}}
\label{tab:visual_quality}
% \vspace{-4mm}
\end{table}

\subsubsection{\textbf{User Survey}} 
\label{sec:human_quality_assessment}

To investigate how well humans can identify deepfakes in MultiFakeVerse, we conducted a user study with 18 participants.\footnote{All procedures in this study were conducted in accordance with Monash University Human Research Ethics Committee approval for Project ID 47707.} We selected 50 random samples (25 real and 25 fake), representing a range of modification levels. Each participant was asked to classify the images as real or fake and for the images they identified as fake, to specify the manipulation level(s). The human accuracy of classifying an image as real or fake came out as 61.67\%, i.e. more than one-thirds of the times, the users tend to misclassify the images. The class-wise average F1 score was 61.14\%. Analyzing the human predictions of manipulation levels for the fake images, the average intersection over union between the predicted and actual manipulation levels was found to be 24.96\%. This shows that it is non-trivial for human observers to identify the regions of manipulations in our dataset.
% The user study results presented in Table~\ref{tab:user_study} indicate that the deepfake content in MultiFakeVerse dataset is very challenging to detect for humans.

% \begin{table}[t]
% \centering
% \caption{\textbf{User study results for MultiFakeVerse and SIDA.}}
% \vspace{-2mm}
% \scalebox{0.85}{
% \begin{tabular}{l|cccc}
% \toprule[0.4mm]
% \rowcolor{mygray} \textbf{User Study} & \textbf{Acc.} & \textbf{AP@0.1} & \textbf{AP@0.5} & \textbf{AR@1} \\ \hline \hline
% SIDA &  &  &  &  \\
% MultiFakeVerse & 61.67\% &  &  &  \\
% \bottomrule[0.4mm]
% \end{tabular}}
% \label{tab:user_study}
% % \vspace{-4mm}
% \end{table}

\subsubsection{\textbf{Computational Cost}}
A total of 845,286 API calls were made to both Gemini and GPT, respectively, for edit suggestions, resulting in an overall cost of $\sim$1000 USD.
For image generation through Gemini API, the total cost was 2867 USD. The cost for the GPT-Image-1 model based images was 200 USD and ICEdit images were generated on one Nvidia A6000 GPU in 24 hours.

\begin{table*}[t]
\centering
\caption{Benchmarking of MultiFakeVerse dataset on Deepfake detection in zero-shot and supervised finetuning setting. The numbers in bracket indicate the changes after supervised finetuning.}
\label{tab:quantitative}
\scalebox{0.8}{
\begin{tabular}{l|c||cc|cc|cc}
\toprule[0.4mm]
\rowcolor{mygray} 
\textbf{Method}  &  \textbf{Year} & \multicolumn{2}{c|}{\textbf{Real}}& \multicolumn{2}{c|}{\textbf{Fake}}& \multicolumn{2}{c}{\textbf{Overall}} \\
\rowcolor{mygray} &  & Acc ($\uparrow$) & F1 ($\uparrow$) & Acc ($\uparrow$) & F1 ($\uparrow$) & Acc ($\uparrow$) & F1 ($\uparrow$)\\ \hline \hline
% \multirow{7}{*}{Zero-Shot}& FakeVLM~\cite{}&  &  &  &  & &  & & &  &\\       
CnnSpot~\cite{wang2020cnn} & 2021 & 100.00 (6.16$\downarrow$) & 19.00 (69.88$\uparrow$) & 0.00 (97.97$\uparrow$) & 0.00 (98.62$\uparrow$) & 50.02 (45.88$\uparrow$) & 9.54 (84.21$\uparrow$)\\ 
% & LGrad~\cite{} & 2023 &  &  &  & &  & & &  &\\ 
% & FakeShield~\cite{}&  &  &  &  & &  & & &  &\\
Trufor~\cite{Guillaro_2023_CVPR}& 2023 & 84.00 & 18.52 & 15.27 & 26.07 & 49.64 & 22.30\\
AntiFakePrompt~\cite{chang2024antifakepromptprompt} & 2024 & 63.30 & 30.47 & 70.45 & 80.63 & 66.87 & 55.55\\
% & SIDA-7B & 2024 &  &  &  & &  & & &  &\\ 
SIDA-13B~\cite{huang2024sida} & 2024 & 67.40(23.07$\uparrow$) & 21.30(64.09$\uparrow$) & 44.55(52.94$\uparrow$) & 60.08(38.09$\uparrow$) & 55.97(38.01$\uparrow$) & 40.69(51.09$\uparrow$)\\ 
% \multirow{7}{*}{Supervised}& FakeVLM~\cite{}&  &  &  &  & &  & & &  &\\ 

% \multirow{4}{*}{Supervised}& AntiFakePrompt~\cite{} & 2024 &  &  &  & &  & & &  &\\      
% & CnnSpot~\cite{} & 2021 &  &  &  & &  & & &  &\\ 
% % & LGrad~\cite{} & 2023 &  &  &  & &  & & &  &\\ 
% % & FakeShield~\cite{}&  &  &  &  & &  & & &  &\\ 
% % & SIDA-7B & 2024 &  &  &  & &  & & &  &\\
% & Trufor~\cite{Guillaro_2023_CVPR}& 2023 &  &  &  &  &  &  & &  &\\
% & SIDA-13B & 2024 &  &  &  & &  & & &  &\\ 

\bottomrule[0.4mm]
\end{tabular}}
\end{table*}

\section{Benchmarks and Experiments}
This section outlines the benchmark protocol for MultiFakeVerse along with the used evaluation metrics. The goal is to detect and localize image based manipulations.

\subsubsection{\textbf{Data Partitioning and Evaluation Metrics}}
We split the dataset into \textit{train}, \textit{validation}, and \textit{test} sets as follows: first we randomly select 70\% of the real images as train set, 10\% real images as validation set and remaining 20\% as test set. Next we add the images generated corresponding to each of the real images in the same set as the real image. We evaluate detection using image-level accuracy and F1 scores. For forgery localization, we use Area Under the Curve (AUC), F1 scores
and Intersection over Union (IoU).

% We also prepared subsets for person level, Person-object interaction level, Person-scene interaction level (See Table~\ref{tab:dataset_stats}).

% \subsection{Implementation Details}
% For benchmarking image based deepfake localization, we consider two state-of-the-art .

% Since these two methods require pre-trained features, we extracted
% the SOTA features from \sg{************}, which are the SOTA methods specifically designed for image based deepfake localization. We followed the original
% settings for *******. For ******, we implemented it using the

% FreDect~\cite{}, Fusing~\cite{}, Gram-Net~\cite{}, UnivFD~\cite{}, LGrad~\cite{}, and LNP~\cite{}
\subsubsection{\textbf{Detection Evaluation}} We compare MultiFakeVerse against several SOTA deepfake detection methods on the test set, including CnnSpot~\cite{wang2020cnn}, AntifakePrompt~\cite{chang2024antifakepromptprompt}, TruFor~\cite{Guillaro_2023_CVPR} and VLM based SIDA~\cite{huang2024sida}. To ensure a fair comparison, we first evaluate these models on our dataset using their original pre-trained weights (zero-shot setting), and then retrain two of them (CNNSpot and SIDA) with our train dataset to assess performance improvements. Table~\ref{tab:quantitative} demonstrates that these models trained on earlier inpainting-based fakes struggle to identify our VLM-Editing based forgeries, particularly, CNNSpot tends to classify almost all the images as real. AntiFakePrompt has the best zero-shot performance with 66.87\% average class-wise accuracy and 55.55\% F1 score. After finetuning on our train set, we observe a performance improvement in both CNNSpot and SIDA-13B, with CNNSpot surpassing SIDA-13B in terms of both average class-wise accuracy (by 1.92\%) as well as F1-Score (by 1.97\%).\\
% \sg{PSCC-Net~\cite{}, MVSS-Net~\cite{}, and HIFI-Net~\cite{}}
\subsubsection{\textbf{Localization Evaluation}} We evaluate the localization performance of SIDA-13B~\cite{huang2024sida} model on our test set, using the perturbation masks obtained in Section \ref{subsec:manipulation_analysis}, both using their pre-trained weights as well as after finetuning on our dataset. In zero-shot setting, the model has an IoU of 13.10, F1-Score 19.92 and AUC 14.06. After finetuning on our train dataset, the localization performance of the SIDA-13B model remains at an IoU of 24.74, F1-Score 39.40 and AUC 37.53, signifying that the model struggles to accurately identify the manipulated image regions. Thus, there is room for further improvement in the forgery localization methods, with MultiFakeVerse as an ideal benchmark.
% The IoU score changes from  to 24.74, F1-Score from  to 39.40 and AUC from  to 37.53.

% \noindent \textbf{Benchmark Comparison.} We conducted additional experiments (Table~\cite{}) to compare the performance on deepfake region localization and classification on MultiFakeVerse and SID-Set~\cite{}. There is a significant drop in ****** for localization performance as compared to SIDA. A similar pattern is also observed for ********

\section{Conclusion}
This paper introduces MultiFakeVerse, the largest image-based dataset for spatial deepfake localization. A thorough benchmarking using SOTA detection and localization methods reveals that VLM-based manipulations are predominantly unidentifiable for both humans as well as detection models solely trained on the existing inpainting-based datasets; underscoring the increased complexity and realism of MultiFakeVerse. These results demonstrate that the proposed dataset is a valuable resource for advancing the development of next-generation deepfake localization techniques.

\noindent \textbf{Limitation.} Like other deepfake datasets, MultiFakeVerse presents an imbalance between the number of real and fake images, which may affect training dynamics and evaluation outcomes.

\noindent \textbf{Broader Impact} With its diverse and content-rich manipulations, MultiFakeVerse is expected to foster the development of more robust and generalizable deepfake detection and localization models beyond facial and object manipulation, especially in image-based settings that reflect real-world usage scenarios.

\noindent \textbf{Ethics statement} We acknowledge potential ethical concerns related to the misuse of manipulated images, such as unauthorized use of identities or image-based misinformation. To mitigate these risks, MultiFakeVerse is distributed under a strict research-only license. Access to the dataset requires agreement to an end-user license agreement (EULA) that limits usage to non-commercial, academic research.

\bibliographystyle{ACM-Reference-Format}
\bibliography{ref}

% \appendix









\end{document}